\documentclass[english,pss,pssb]{w-art}
\usepackage{times}
\usepackage[T1]{fontenc}
\usepackage[latin1]{inputenc}
\usepackage{amsmath}
\usepackage{graphicx}
\usepackage{amssymb}

\makeatletter
\usepackage{cite}

\usepackage{babel}
\makeatother
\begin{document}

\title[
\title{Coupled nanopillar waveguides}\maketitle
]{Coupled nanopillar waveguides:\\
optical properties and applications}

\author[D. N. Chigrin]{Dmitry N. Chigrin\footnote{Corresponding author: e-mail: {\sf chigrin@th.physik.uni-bonn.de}, Phone: +49 228 73 2046, Fax: +49 228 73 3223}\inst{1}}

\author[S. V. Zhukovsky]{Sergei V. Zhukovsky\inst{1}}

\author[A. V. Lavrinenko]{Andrei V. Lavrinenko\inst{2}}

\author[J. Kroha]{Johann Kroha\inst{1}}

\address{\inst{1} Physikalisches Institut, Universit\"at Bonn, Nussallee
12, D-53115 Bonn, Germany}

\address{\inst{2} COM-DTU, Department of Communications, Optics and Materials,
NanoDTU, Technical University of Denmark, Building 345V, DK-2800 Kgs.
Lyngby, Denmark}

\keywords{Coupled and nanopillar waveguides, periodic and aperiodic structures,
photonics devices}

\subjclass[pacs]{42.60.Fc, 42.55.Tv, 42.82.Gw.}

\begin{abstract}
In this paper we review basic properties of coupled periodic and aperiodic
nanopillar waveguides. A coupled nanopillar waveguide consists of
several rows of periodically or aperiodically placed dielectric rods
(pillars). In such a waveguide, light confinement is due to the total
internal reflection, while guided modes dispersion is strongly affected
by the waveguide structure. We present a systematic analysis of the
optical properties of coupled nanopillar waveguides and discuss their
possible applications for integrated optics.
\end{abstract}
\maketitle

\section{Introduction}

\begin{figure}[t]
\begin{centering}\includegraphics[width=0.9\columnwidth]{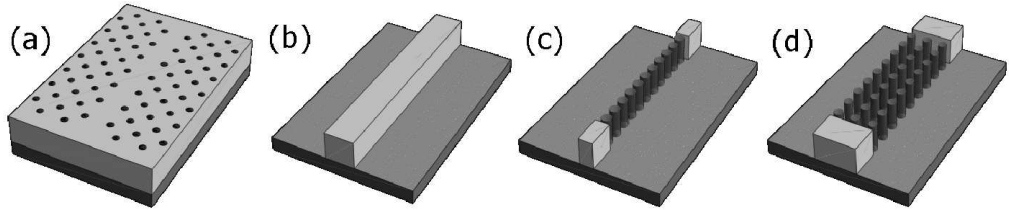}\par\end{centering}

\caption{\label{Fig:Sketch}Optical waveguides: (a) photonic crystal waveguide,
(b) dielectric waveguide, (c) nanopillar waveguide, and (d) coupled
nanopillar waveguide.}
\end{figure}

Photonic crystals (PhCs) are known for offering unique opportunities for
controling the flow of light by acting as waveguides, cavities, dispersive
elements, etc \cite{ref1,ref2,ref3,ref4}. Photonic crystal waveguides
(PCW) are one of the promising examples of PhCs applications at micron
and sub-micron length-scales. They can be formed by removing one or
several lines of scatterers from the PhC lattice (Fig.~\ref{Fig:Sketch}a).
PCW based on PhCs with different two-dimensional (2D) lattices of
both air holes in a dielectric background and dielectric rods in air
were reported \cite{ref1,ref2,ref3}. Light confinement in PCW is
obtained due to a complete photonic bandgap (PBG), in contrast to
the standard guiding mechanism in a conventional dielectric waveguide
(Fig.~\ref{Fig:Sketch}b). It was theoretically predicted that a
PhC waveguide can possess loss-free propagation as soon as a guiding
mode falls into a complete PBG. However, progress in PhC research
has revealed that losses are inevitable and sometimes might be
rather high even in spite of broad PBG. Special optimization efforts
are now intensively applied for decreasing optical losses and the results
are quite promising \cite{ref5,ref6}.

At the same time, PBG guiding is not the only waveguiding mechanism
in a PhC. Unique anisotropy of PhCs can cancel out the natural diffraction
of the light, leading to the self-guiding of a beam in a non-channel
PCW \cite{ref7,ref8,ref9}. The common principle of index guiding
(guiding due to total internal reflection) can be also found in periodic
systems. It is rather straightforward if a waveguide is organized
as a defect in a lattice of holes in a dielectric material. Then, the
channel itself has higher index of refraction than the average index of 
the drilled or etched medium. Topologically inverted systems like periodic
arrays of rods or nanopillars placed in air can also provide waveguiding
due to index difference \cite{ref2}. However, fabrication of rod
arrays on the nanoscale is a relatively difficult technological
problem.

The recent progress in the fabrication of alternative structures, namely 
nanopillar waveguides (NPW),
has proved the relevance of their study not only as a useful theoretical
model. For example, two-dimensional (2D) silicon-on-insulator (SOI) pillar PhC 
have recently been
fabricated and characterized \cite{ref10}. Sandwich-like
structures have also been successfully realized in $GaAs/Al_{x}O_{y}$
material system \cite{ref11}. Membrane-like structures have been realized,
based on polymer membranes incorporating $Si$ rods \cite{ref12}.
Recently, various combinations of active materials inserted in single
nanowires or arrays of nanopillars have been under attention as well
\cite{ref13}.
It is important to point out that all of the above mentioned studies do not
only present a successful practical realization of the pillar PhC
structures, but also report transmission efficiencies and out-of-plane
radiation losses comparable with the 2D PhC based on hole geometry.

A one-dimensional (1D) chain of rods placed at equal distance from
one another (Fig.~\ref{Fig:Sketch}c) possesses guiding properties
as was shown by Fan et al. \cite{ref2}. The fundamental mode of such
a \textit{\emph{periodic nanopillar waveguide}} lies below the light
line and below the first PBG corresponding to the 2D PhCs with a square
lattice of the same rods. Guiding is due to total internal reflection.
A better confinement of light can be achieved, if several 1D periodic
chains are placed in parallel (Fig.~\ref{Fig:Sketch}d) \cite{ref14}.
Such waveguides are called \textit{coupled
nanopillar waveguides} (\textit{CNPWs}) and are designated as W\textit{n},
where \textit{n} is the number of parallel rows comprising the CNPW.
In building a CNPW both the longitudinal and the transverse relative shift
between individual waveguides can be arbitrary, and thus, a high flexibility
in dispersion engineering can be achieved.

In this paper, we review basic properties of coupled nanopillar waveguides
and discuss their possible applications for integrated optics. In
Section~\ref{Sec:Dispersion}, a CNPW is introduced and possible
ways to tune the CNPW dispersion are discussed. The transmission efficiency
of 2D and 3D CNPWs is reported in Section~\ref{Sec:Transmission}.
The route to improve the coupling between a nanopillar waveguide and an
external dielectric waveguide (like an optical fiber) 
is discussed in Section~\ref{Sec:QNPW}
with respect to aperiodic NPWs. Possible applications of coupled periodic
and aperiodic nanopillar waveguides are discussed in 
Section~\ref{Sec:Applications}.
Section~\ref{Sec:Conclusion} concludes the paper.

\section{\label{Sec:Dispersion}Dispersion engineering}

\subsection{Dispersion tuning}

\begin{figure}[t]
\begin{centering}
\includegraphics[width=0.8\columnwidth]{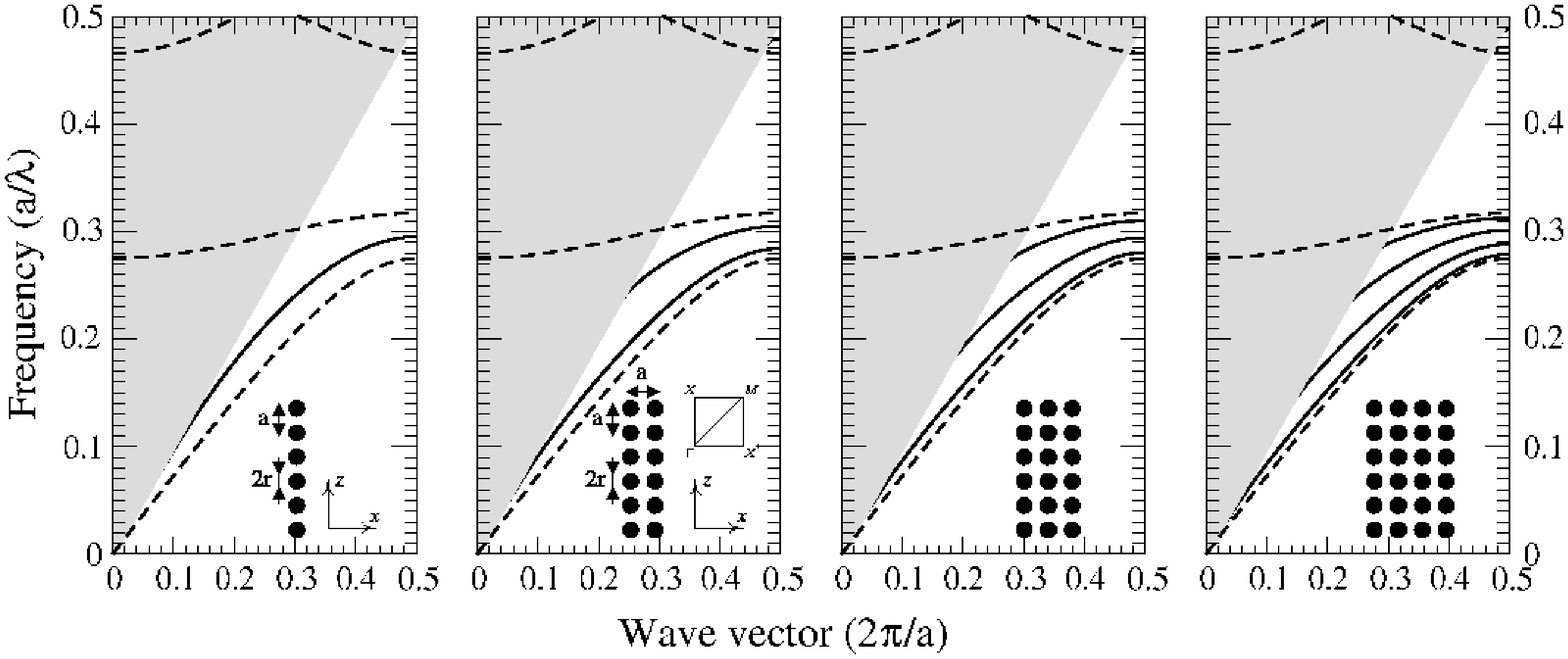}\par\end{centering}

\caption{\label{Fig:DisSquare}Dispersion diagrams for CNPWs with 1, 2, 3
and 4 rows. The insets show a sketch of the waveguides. The coordinate
system together with the first quadrant of the first Brillouin zone
of the square lattice with period $a$ are also shown. The grey areas depict
teh continuum of radiated modes lying above the light line. Guided modes
are shown as black solid lines. The projected band structure of the infinite
2D PhC is shown as dashed lines. Here $\varepsilon=13.0$ and \textit{r}
= $0.15a$.}
\end{figure}

In reference \cite{ref2} it was shown that a single row of periodically
placed dielectric rods is effectively a single-mode waveguide within
a wide frequency range (Fig.~\ref{Fig:DisSquare}, left panel). It
has a well confined fundamental mode. Attaching one, two or more identical
W1 waveguides in parallel to the original one produces a coupled-waveguide
structure \cite{ref14}. It is well known in optoelectronics that
this leads to the splitting of the original
mode into \textit{n} modes, where \textit{n} is the number of coupled
waveguides \cite{ref15}.

In Fig.~\ref{Fig:DisSquare}, dispersion diagrams for W1, W2, W3
and W4 CNPWs are shown. All rods are placed at the vertices of a
square lattice. To model a CNPW dispersion we used the plane-wave expansion
method (PWM) \cite{ref16}. The supercell consists of one
period in the \textit{z} direction and 20 periods in the \textit{x}
direction, where \textit{n} periods occupied by dielectric rods were
placed in the center of the supercell. The waveguide is oriented 
along the \emph{z}-axis (Fig.~\ref{Fig:DisSquare}). 
The calculations were performed for 2D structures and for TM polarization. 
The $n$ modes of the CNPW are bound between the $\Gamma-X$
and $X-M$ projected bands of the corresponding infinite PhC of a
2D square lattice of rods (Fig.~\ref{Fig:DisSquare}, dashed lines)
\cite{ref14}. All modes are effectively localized 
within the waveguide region. Near the irreducible Brillouin zone (IBZ) 
boundary the dispersion is strongly affected by the system periodicity.

\begin{figure}[b]
\begin{centering}\includegraphics[width=0.55\columnwidth]{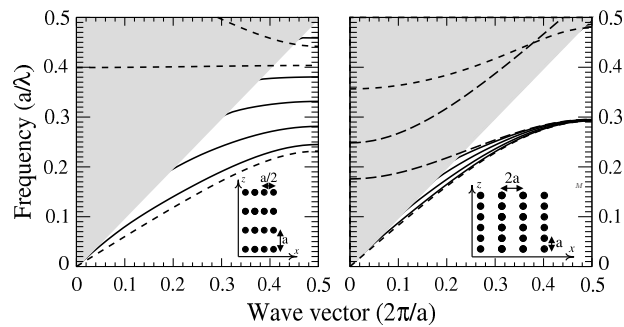}\par\end{centering}

\caption{\label{Fig:DisRec}The same as in Fig.~\ref{Fig:DisSquare} for
CNPWs with different transverse offsets, $d=0.5a$ (left) and $d=2.0a$
(right). The insets show a sketch of waveguides and coordinate system.}
\end{figure}

It is well known that by varying the filling factor, i.e. the rod radius,
and the dielectric constant of the rods, one can tailor the frequency range
and slope of the PhC bands. Taking into account that the CNPW modes
are bound by $\Gamma-X$ and $X-M$ bands of the corresponding infinite
2D PhC, a proper frequency adjustment of nanopillar waveguide modes
can be done by changing these two parameters. Decreasing the dielectric
constant of the rods, while keeping their radius constant, pushes the bundle
of \textit{n} CNPW modes to higher frequencies. The modes shift towards
lower frequencies, if the nanopillar radius increases, with fixed
dielectric constant. In general, the mode tuning follows the
rule: the larger the average refractive index of the system, the lower the
mode frequencies \cite{ref14}.

Another option for tuning the mode dispersion of CNPW is to change the distance
between individual waveguides, the transverse offset. Examples are
shown in figure~\ref{Fig:DisRec} for two transverse offsets, $d=0.5a$
(left) and $d=2.0a$ (right). In these cases the rods are situated at
the vertices of a rectangular lattice. While the mode overlap of individual
waveguides is larger (smaller) for close (far) positioned waveguides,
the coupling strength is stronger (weaker). For two identical waveguides,
this in turn results in stronger (weaker) mode splitting, 
$\beta_{\pm}=\mathit{\beta}\pm\mathit{\kappa}$, with respect to the 
propagation constant $\beta$ of the uncoupled NPW.
Here \emph{$\kappa$} is a coupling coefficient \emph{}\cite{ref15}.
Note, that the CNPW mode frequencies are still bounded by the position of the
projected band structure of the corresponding infinite rectangular 
PhC (Fig.~\ref{Fig:DisRec}, dashed lines).

\begin{figure}[t]
\begin{centering}\includegraphics[width=0.75\columnwidth]{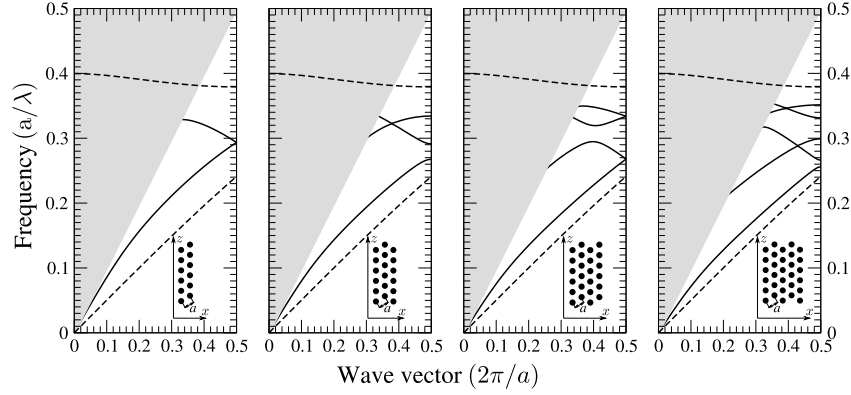}\par\end{centering}

\caption{\label{Fig:DisTr}The same as in Fig.~\ref{Fig:DisSquare} for triangular
lattices W2, W3, W4 and W5 waveguides. The insets show a sketch of the 
waveguides and the coordinate system. Here $\varepsilon=13.0$ and $r=0.26a$.}
\end{figure}

The last parameter which may affect the dispersion of a CNPW is the
longitudinal shift between its individual rows. In figure~\ref{Fig:DisTr}
the dispersion diagrams for CNPW with rods placed in the vertices
of a triangular lattice are shown for W2, W3, W4 and W5 waveguides.
The orientation of the waveguides coincides with the $\Gamma-X$ direction
of the triangular lattice. The mode splitting in a ``triangular lattice''
W2 waveguide strongly depends on the propagation constant 
(Fig.~\ref{Fig:DisTr},
left panel), being large for small $\beta$ and vanishing near the
IBZ boundary. This is in contrast to a ``square lattice'' W2 waveguide
(Fig.~\ref{Fig:DisSquare}), where the mode splitting is approximately
constant for all propagation constants. The mode multiplicity near the
IBZ boundary leads to regions with negative dispersion (backward propagating
waves) of the second mode. For CNPWs with the number of rods larger than 
two (Fig.~\ref{Fig:DisTr}, right panels) it results, furthermore, 
in the formation of mini-bandgaps
and multiple backward waves regions in the dispersion.
Note that in spite of the complex nature of the mode splitting,
CNPW modes are still bounded by the projected bands of the corresponding
triangular lattice PhC.

\begin{figure}[b]
\begin{centering}\includegraphics[width=0.7\columnwidth]{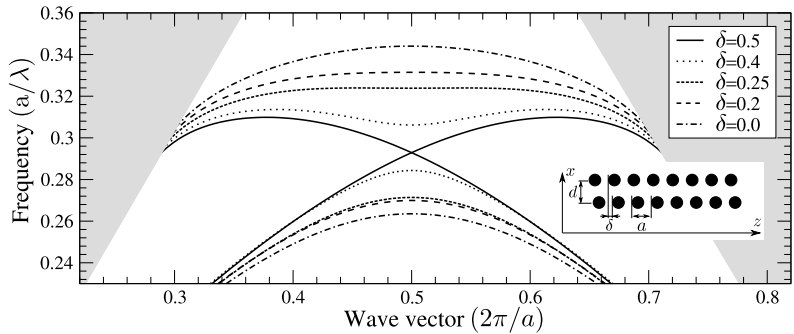}\par\end{centering}

\caption{\label{Fig:shift}Dispersion diagrams of the W2 CNPW for different
longitudinal shifts \textit{$\delta$}. The insets show a sketch of two
parallel periodic waveguides with transverse offset $d$ and longitudinal
shift \textit{$\delta$}. Here $\varepsilon=13.0$, $r=0.15a$ and
$d=a$.}
\end{figure}

The longitudinal shift $\delta$ can be arbitrarily set to any value
between $\delta=0$ and $\delta=0.5a$. The concomitant dramatic changes
in the CNPW dispersion are illustrated in figure~\ref{Fig:shift}
for the case of W2 waveguide. Starting from the simple mode splitting
for $\delta=0$ one can have a very flat second band for $\delta\approx0.25a$,
with negative dispersion regions in the second band for $\delta>0.25a$
and degenerate first and second bands at the IBZ boundary for $\delta=0.5a$\textit{.}
By combining such shifted W2 waveguides and appropriately chosing the 
rod radius and transverse
offset one obtains large flexibility in designing CNPWs with anomalous
dispersion in the frequency range of interest.

\subsection{Coupled mode model}

To understand qualitatively the physical mechanism of the anomalous dispersion
presented in the last example (Fig.~\ref{Fig:shift}), the coupled
mode theory (CMT) can be used \cite{AVLunpub}. Being an approximate
theory, CMT nevertheless manages to combine a simple physical model
with accurate qualitative and even quantitative results \cite{huang94}.
In what follows, two identical coupled periodic waveguides \textit{a-}W1
and \textit{b}-W1 are arranged in a W2 CNPW. The second waveguide,
\emph{b}-W1, is shifted by \textit{$\delta$} with respect to the
first one (Fig.~\ref{Fig:shift}, inset). We limit ourselves to the
scalar CMT, which in our case corresponds to the TM polarization.

The modes of the W2 waveguide are defined as the solutions of the 2D 
scalar wave equation

\begin{equation}
\left(\frac{\partial^{2}}{\partial x^{2}}+\frac{\partial^{2}}{\partial z^{2}}\right)E\left(x,z\right)+k_{0}^{2}\mathit{\varepsilon}\left(x,z\right)E\left(x,z\right)=0,\label{eq:1}\end{equation}
where the dielectric function of the composite structure is simply the sum
of dielectric functions of the two W1 waveguides, 
$\varepsilon\left(x,z\right)=\mathit{\varepsilon}_{a}\left(x,z\right)+\mathit{\varepsilon}_{b}\left(x,z\right)$.
Here $k_{0}=\omega/c$ is a wave number in vacuum. We are looking
for a solution of Eq.~(\ref{eq:1}) in the form of a linear combination
of the propagating modes of two isolated W1 waveguides \cite{maerz87},
which allows us to separate spatial variables in the form

\begin{equation}
E\left(x,z\right)=\mathit{\Psi}_{a}\left(x\right)\left(f_{a}\left(z\right)e^{-\mathit{i\beta z}}+b_{a}\left(z\right)e^{\mathit{i\beta z}}\right)+\mathit{\Psi}_{b}\left(x\right)e^{-i\beta\delta}\left(f_{b}\left(z\right)e^{-\mathit{i\beta z}}+b_{b}\left(z\right)e^{\mathit{i\beta z}}\right).\label{eq:2}\end{equation}
Here $f_{m}\left(z\right)=F_{m}\left(z\right)\text{exp}\left(i\left(\beta-\mathit{\beta}_{0}\right)z\right)$
and $b_{m}\left(z\right)=B_{m}\left(z\right)\text{exp}\left(-i\left(\beta-\mathit{\beta}_{0}\right)z\right)$
are the slowly varying amplitudes of forward and backward propagating
modes near the Bragg resonance condition of a single periodic W1 waveguide
with period $a$ and $\beta_{0}=\pi/a$. 
The functions $\mathit{\Psi}_{a}\left(x\right)$
and $\mathit{\Psi}_{b}\left(x\right)$ represent the transverse field 
distributions, and indexes $m=a,b$ refer to \textit{a-}W1 and \textit{b}-W1 
waveguides,
respectively. The spatial shift between the two W1 waveguides is accounted
for by the corresponding phase shift $e^{-i\beta\delta}$ of the field
of the \emph{b}-W1 waveguide. Here \textit{$\beta$}
is the propagation constant of a homogenized W1 waveguide of
width $l=2r$ and dielectric constant $\varepsilon_{\mathrm{eff}}\left(x\right)=\left(1/a\right)\int_{z}^{z+a}dz\varepsilon\left(x,z\right)$.
The dependence of the propagation constant $\beta$ on frequency is given
by the standard planar waveguide dispersion relation \cite{ref15}\begin{equation}
\tan^{2}\left(\frac{l}{2}\sqrt{n_{\mathrm{eff}}^{2}k_{0}^{2}-\beta^{2}}\right)=\left(\beta^{2}-k_{0}^{2}\right)\left(n_{\mathrm{eff}}^{2}k_{0}^{2}-\beta^{2}\right)^{-1},\label{eq:3}\end{equation}
where we have introduced the effective index of refraction of the homogenized
waveguide $n_{\mathrm{eff}}=\sqrt{\varepsilon_{\mathrm{eff}}}$. The
transverse field distributions $\mathit{\Psi}_{a}\left(x\right)$
and $\mathit{\Psi}_{b}\left(x\right)$ obey the scalar wave equations

\[
\left(\frac{\partial^{2}}{\partial x^{2}}-\mathit{\beta}^{2}\right)\mathit{\Psi}_{m}\left(x\right)+k_{0}^{2}\mathit{\varepsilon}_{0m}\left(x\right)\mathit{\Psi}_{m}\left(x\right)=0\]
with $m=a,b$ and the transverse dependent dielectric functions $\varepsilon_{0m}\left(x\right)$
being a \emph{z}-average dielectric constant of the m\emph{-th} waveguide\textit{.}
Substituting the mode expansion (\ref{eq:2}) into the scalar wave
equation (\ref{eq:1}) and expanding the dielectric constant $\mathit{\varepsilon}\left(x,z\right)$
in a Fourier series with respect to $z$,

\begin{equation}
\mathit{\varepsilon}\left(x,z\right)=\mathit{\varepsilon}_{a0}\left(x\right)+\mathit{\varepsilon}_{b0}\left(x\right)+\underset{l\neq0}{\sum}\left(\mathit{\varepsilon}_{al}\left(x\right)+\mathit{\varepsilon}_{bl}\left(x\right)\right)e^{-\mathit{il\left(2\pi/a\right)z}},\label{eq:4}\end{equation}
with the Fourier coefficients $\mathit{\varepsilon}_{ml}\left(x\right)$, 
one obtains after some lengthy but straightforward
derivations a system of four ordinary differential equations relating
slowly varying amplitudes of the forward and backward propagating modes
in the two W1 waveguides,

\begin{equation}
\frac{d}{dz}\;\left(\begin{matrix}F_{a}\\
F_{b}\\
B_{a}\\
B_{b}\end{matrix}\right)=i\hat{M}\left(\begin{matrix}F_{a}\\
F_{b}\\
B_{a}\\
B_{b}\end{matrix}\right).\label{eq:5}\end{equation}
 For the propagation constant close to the Bragg point $\beta_{0}=\pi/a$,
the system matrix $\hat{M}$ has the form \begin{equation}
\hat{M}=\left(\begin{array}{llll}
\beta_{0}-\beta & -e^{-i\beta\delta}\kappa_{0} & -\kappa_{0} & 0\\
-e^{i\beta\delta}\kappa_{0} & \beta_{0}-\beta & 0 & -e^{2i\beta_{0}\delta}\kappa_{a}\\
\kappa_{a} & 0 & \beta-\beta_{0} & e^{-i\beta\delta}\kappa_{0}\\
0 & e^{-2i\beta_{0}\delta}\kappa_{a} & e^{i\beta\delta}\kappa_{0} & \beta-\beta_{0}\end{array}\right).\label{eq:6}\end{equation}
To simplify the following analysis, we have kept only two coupling constants,
namely $\kappa_{0}$, accounting for the coupling
between two homogenized waveguides, and $\kappa_{a}$,
describing the waveguide's intrinsic periodic structure. These coupling
constants are defined in a usual way, as overlap integrals of the
transverse field distributions with the corresponding Fourier coefficients
of the dielectric function expansion. The resulting propagation constants
of the supermodes of the W2 waveguide are given as the eigenvalues of the
system matrix $\hat{M}$ (\ref{eq:6})\begin{equation}
\beta_{\mathrm{W2}}\left(\omega\right)=\beta_{0}+\Delta\beta\left(\omega\right),\label{eq:7}\end{equation}
with

\begin{equation}
\Delta\beta\left(\omega\right)=\pm\sqrt{\left(\beta\left(\omega\right)-\beta_{0}\right)^{2}+\kappa_{0}^{2}-\kappa_{a}^{2}\pm\kappa_{0}\sqrt{4\left(\beta\left(\omega\right)-\beta_{0}\right)^{2}-2\kappa_{a}^{2}+2\kappa_{a}^{2}\cos\left(2\pi\delta\right)}}.\label{eq:8}\end{equation}
The implicit dependence of the propagation constant on frequency is given
via the dispersion relation (\ref{eq:3}) of a planar homogenized
waveguide.

\begin{figure}[t]
\begin{centering}\includegraphics[width=0.95\columnwidth]{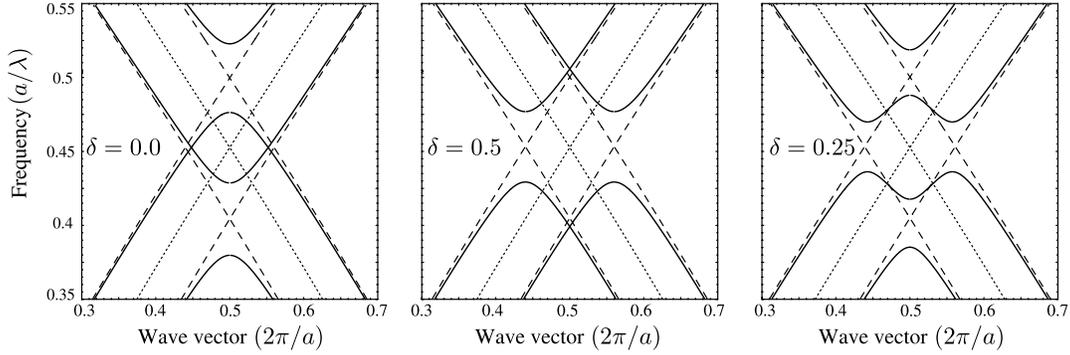}\par\end{centering}

\caption{\label{Fig:cmt}Dispersion diagrams of two coupled periodic waveguides
(solid lines) within the framework of coupled mode theory for three
values of the longitudinal shift $\delta=0.0$ (left), $\delta=0.5$
(center) and $\delta=0.25$ (right). The dotted line is the dispersion
of a homogenized waveguide folded into the first Brillouin zone. 
The dashed lines are the folded dispersions
of two coupled homogenized waveguides. Here $n_{\mathrm{eff}}=1.5$,
$l=0.3a$, $\kappa_{0}=0.06$ and $\kappa_{a}=0.03$.}
\end{figure}

In figure~\ref{Fig:cmt} the dispersion diagram of W2 waveguide calculated
using Eqs.~(\ref{eq:7}-\ref{eq:8}) is presented for three values
of the longitudinal shift $\delta=0.0$ (left), $\delta=0.5$ (center)
and $\delta=0.25$ (right). See figure caption for further details
on the parameters. The dotted line shows the dispersion of a planar homogenized
waveguide calculated using the dispersion relation (\ref{eq:3}) and
folded back into the first Brillouin zone by the Bragg wave vector 
corresponding to the
periodic W1 waveguide. By setting the self-action coupling constant,
$\kappa_{a}$, to zero and choosing some finite value for the inter-row
coupling constant, $\kappa_{0}$, one can reproduce the simple band
splitting within the CMT model. The split modes are shown as dashed lines
(Fig.~\ref{Fig:cmt}).

To analyze the influence of the periodic structure and the longitudinal
shift on the split band structure, we first consider zero longitudinal
shift, $\delta=0.0$. In this situation the detuning of the propagation
constant from the Bragg wave vector, $\beta_{0}$, is given by $\Delta\beta=\pm\sqrt{\left(\Delta\pm\kappa_{0}\right)^{2}-\kappa_{a}^{2}}$,
where $\Delta=\left(\beta-\beta_{0}\right)$ is the detuning of the
propagation constant of the homogenized waveguide from the Bragg point.
The propagation factor of the supermodes is given by the exponential $e^{\pm i\beta_{0}z}e^{\pm i\left(\sqrt{\left(\Delta\pm\kappa_{0}\right)^{2}-\kappa_{a}^{2}}\right)z}$,
which corresponds to propagating modes only if $\left(\Delta\pm\kappa_{0}\right)^{2}-\kappa_{a}^{2}>1$.
In the opposite situation, there are two bandgaps at Bragg wave vector
$\beta_{0}$ with central frequencies corresponding to $\Delta=\pm\kappa_{0}$.
These bandgaps are due to the destructive interference of the first
forward propagating and the first backward propagating supermodes
and the second forward propagating and the second backward propagating
supermodes, respectively, as can be seen from the left panel of figure~\ref{Fig:cmt}.
In the case of half-period shifted W1 waveguides, $\delta=0.5$, the
detuning of the propagation constant and the supermode propagation
constant are given by $\Delta\beta=\pm\left(\sqrt{\Delta^{2}-\kappa_{a}^{2}}\pm\kappa_{0}\right)$
and $e^{\pm i\left(\beta_{0}\pm\kappa_{0}\right)z}e^{\pm i\left(\sqrt{\Delta^{2}-\kappa_{a}^{2}}\right)z}$.
In this case two bandgaps exist at shifted Bragg wave vectors $\beta_{0}\pm\kappa_{0}$
with central frequency at $\Delta=0.0$. This corresponds to the destructive
interference of the first forward propagating and the second backward
propagating supermodes and vice versa (Fig.~\ref{Fig:cmt}, center).
It is important to mention here that at the Bragg condition, $\beta_{0}$,
(IBZ boundary) the first forward propagating and the first backward
propagating supermodes are in phase, which leads to the degeneracy
of the first and second bands at the IBZ boundary (Fig.~\ref{Fig:DisTr}).
The shift of the Bragg condition away from the IBZ 
boundary, $\beta_{0}\pm\kappa_{0}$,
is a reason for the appearance of a region with negative dispersion in the
second band (Fig.~\ref{Fig:DisTr}). In the case of arbitrary shift
between the W1 waveguides a destructive interference takes place between
all possible pair combinations of forward and backward propagating
supermodes leading to the formation of four bandgaps and anomalous
dispersion (Fig.~\ref{Fig:shift}). In the right panel of figure~\ref{Fig:cmt}
an example of a CMT dispersion diagram is shown for the case of quarter-period
shifted waveguides, $\delta=0.25$.

\section{\label{Sec:Transmission}Transmission efficiency }

\begin{figure}[t]
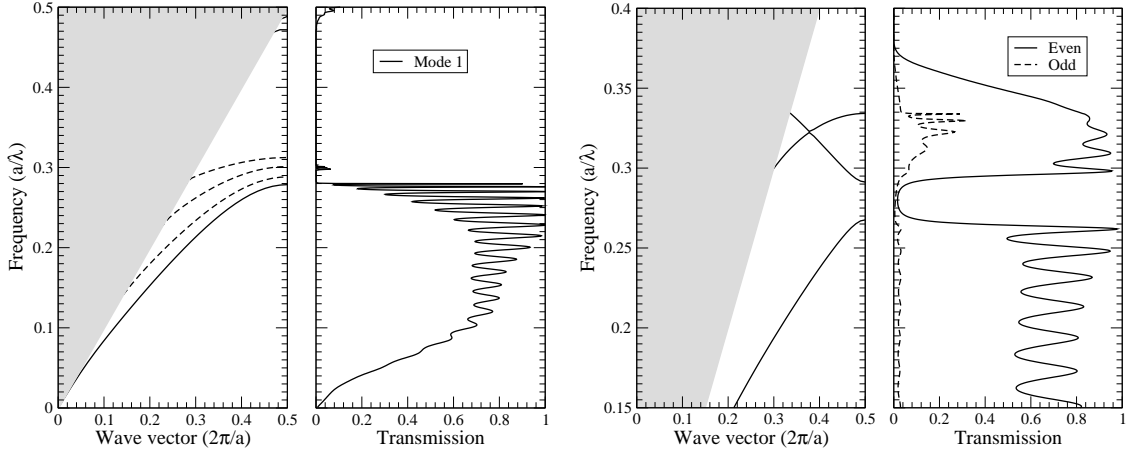

\begin{centering}\includegraphics[clip,width=0.485\columnwidth]{Figs/fig7a}\hfill{}\includegraphics[clip,width=0.49\columnwidth]{Figs/fig7b}\par\end{centering}

\caption{\label{Fig:trans2D}Left: Dispersion diagram and transmission spectra 
of a 2D {}``square-lattice'' W4 CNPW. The fundamental mode was excited.
Here $\varepsilon=13.0$ and $r=0.15a$. Right: Dispersion diagram and 
transmission spectra of a 2D W3 {}``triangular-lattice'' CNPW.
Solid line -- even excitation, dashed line -- odd excitation. Here
$\varepsilon=13.0$ and \textit{$r=0.26a$.}}
\end{figure}

An important characteristic of the novel waveguides is their transmission
efficiency. To analyze transmission efficiencies of different 2D and
3D CNPW, the finite difference time domain (FDTD) method \cite{ref18}
with perfectly matched layers as absorbing boundary conditions at
all sides and a resolution of 16 grid points per lattice constant
is used here. The modes are excited by a Gaussian-shaped temporal impulse,
the Fourier transform of which is broad enough to cover the frequency
range of interest. Fields are monitored by input and output detectors.
The transmitted wave intensities are normalized by the ones of the incident
waves.

The calculated transmission spectrum of a 20 periods long, straight 
``square-lattice''
W4 CNPW is shown in the left panels of Fig.~\ref{Fig:trans2D}. There
are four modes under the light line as it is shown in the band diagram.
In Fig.~\ref{Fig:trans2D} the transmission of the fundamental mode is 
shown directly together with the dispersion diagram. The W4 waveguide displays
high transmission efficiency (close to 100\%) over a broad spectral
range. The position of the cut-off frequency is clearly seen in the
spectrum.

\begin{figure}[t]
\begin{centering}\includegraphics[clip,width=0.45\columnwidth]{Figs/fig8a}\hfill{}\includegraphics[width=0.38\columnwidth]{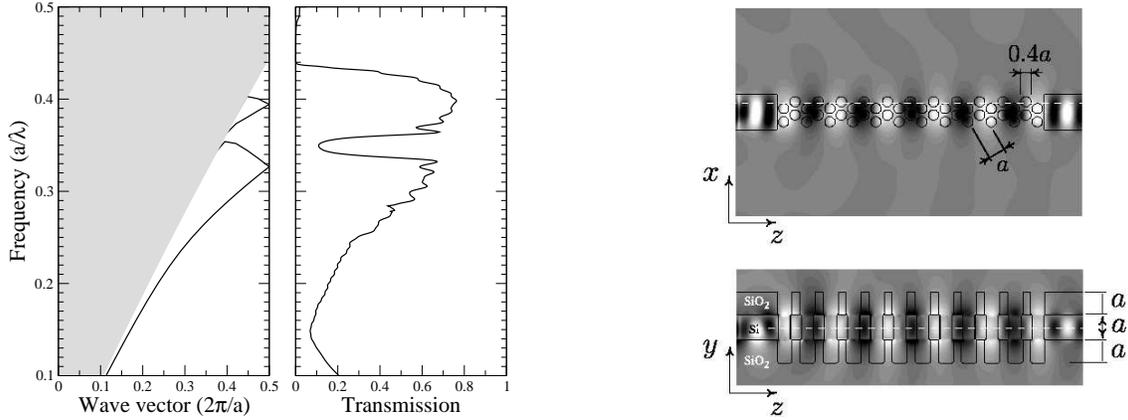}\par\end{centering}

\caption{\label{Fig:trans3D}(Left) Transmission spectra of even mode and
dispersion diagram of 3D W4 SOI {}``triangular lattice'' CNPW. (Right)
Field distribution inside W4 SOI CNPW in horizontal (top) and vertical
(bottom) planes. Grey levels mark electric field amplitude. Black
contours correspond to waveguide structure. White dashed lines depict
positions of the corresponding cuts. Here \textit{\emph{radius is}}
\textit{$r=0.2a$,} the total nanopillar height is $h=3a$, the thickness
of $Si$ layer equals to $a$. Dielectric constants of $Si$ and $SiO_{2}$
were chosen as $\varepsilon=11.5$ and $\varepsilon=2.1025$, respectively.}
\end{figure}

In the right panels of figure~\ref{Fig:trans2D}, the band structure
and transmission spectra are shown for the W3 {}``triangular-lattice''
CNPW. A 20 period long, straight CNPW is cut in the $\Gamma-X$ direction
of the triangular lattice. A substantial suppression of the transmission 
is seen in
the spectrum, coinciding exactly with the position of the mini bandgap
in the band structure. Changing the parity of the signal field distorts
the spectrum reflecting the mode symmetries. The even mode displays high
transmission efficiency (close to 100\%) over a broad spectral range.
The odd mode has a lower level of transmission and is mostly transmitted at
higher frequencies. Here, by odd and even modes we understand the corresponding
first two fundamental modes of a conventional dielectric waveguide.
The surprisingly high transmission of the even mode above the cutoff frequency,
$\omega \approx 0.34$, can be explained by the resonant behavior of the folded
radiation mode with negative group velocity \cite{ref19}. We found
similar behavior above cutoff for other {}``triangular-lattice''
CNPW structures.

An example of 3D calculations for an SOI W4 {}``triangular-lattice''
CNPW are presented in Fig.~\ref{Fig:trans3D}. In the left panel the
dispersion diagram of the structure are presented, while its transmission
spectrum for the even mode is plotted in the central panel. The dispersion
diagram was calculated using 3D supercell PWM. In general, the transmission
spectrum is very similar to the corresponding spectrum of 2D structure
(Fig.~\ref{Fig:trans2D}, right). The transmission band is rather broad
with 80\% transmission efficiency at maximum and a sizable stopband
at the mini bandgap frequencies. We attribute the moderate level of 
transmission
to the impedance mismatch at the conventional waveguide--nanopillar
waveguide interface. In the right panel of Fig.~\ref{Fig:trans3D},
the steady-state electric field distribution inside the CNPW is shown for
a monochromatic source with normalized frequency $\omega=a/\lambda=0.3$.
The field is well confined within the waveguide core in both the horizontal
and vertical planes. there is no evidence for strong energy leakage 
into the substrate.

\section{\label{Sec:QNPW}Aperiodic nanopillar waveguides}

In addition to allowing arbitrary variation of the period and displacement
(which is one of the advantages of the nanopillar waveguides as opposed
to the PCWs), CNPWs allow arbitrary modification of the longitudinal
geometry. A localized change of the properties introduced in one or
several nanopillars would create a point defect, which functions as
a resonator \cite{ref2,Johnson}. The design of such micro-resonators
on the scale of a few wavelengths is essential for integrated optics
applications. Ideally, such resonators should combine the apparently
contradictory features of a high \emph{Q}-factor and of a sufficiently
good coupling to a waveguide terminal to inject or extract light into
or from the resonator. Due to the absence of a complete bandgap, the
breaking of translational symmetry inevitably results in radiation
losses of the resonator mode, which raises the need for optimizing
the \emph{Q}-factor of the resonator in 1D nanopillar waveguides.
There have been some proposals to decrease the losses based on either
mode delocalization \cite{Deloc1} or on the effect of multipole cancellation
\cite{Johnson}. A delocalized mode typically suffers from a decrease
of the \emph{Q}-factor. On the other hand, the spatial radiation loss
profile of a mode described in Ref.~\cite{Johnson} has a nodal line along
the waveguide axis, which means poor coupling to any components coaxial
with the waveguide.

Other than by means of a point defect, a resonant system can also
be created by changing the periodic arrangement of nanopillars into a
non-periodic one. We show that the use of such aperiodically ordered
waveguide leads to improved coupling to the coaxial terminal without
considerably sacrificing the \emph{Q}-factor of the resonant modes.
We use fractal Cantor-like NWPs as an example \cite{zhukJOSA}. To
construct aperiodic NPW, nanopillars of equal radius are arranged in a 
1D chain, where the distances between adjacent pillars are given
by the Cantor sequence. If we denote~$S$ and~$L$ for short and
long distance ($d_{S}$~and~$d_{L}$), respectively, the Cantor
sequence is created by the inflation rule $L\to LSL,\quad S\to SSS$
and unfolds in the following self-similar fashion, which represents
a series of middle third Cantor prefractals $L\to LSL\to LSL\, SSS\, LSL\to LSLSSSLSL\, SSSSSSSSS\, LSLSSSLSL\to\cdots.$

\begin{figure}
\begin{centering}\includegraphics[width=0.95\columnwidth]{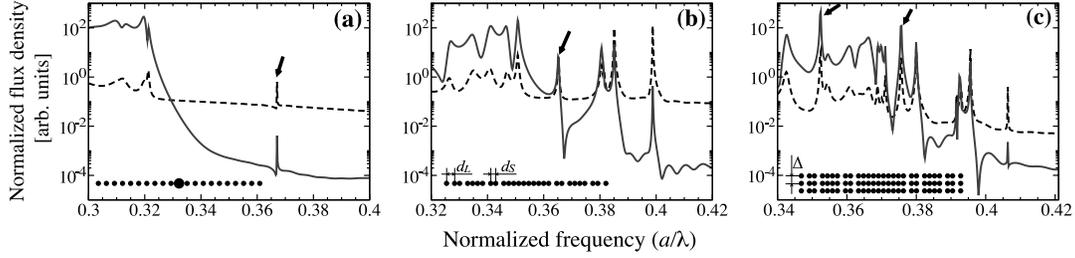}\par\end{centering}

\caption{Normalized energy flux of electromagnetic radiation escaping from the
resonator into the terminal (solid line) and elsewhere (dashed
line) for three nanopillar structures shown in the insets: a W1 with
a point defect \cite{Johnson} (left); a W1 with Cantor-like longitudinal
geometry (center) and a W3 Cantor-like CNPW (right). Here, $d_{S}=0.5a$,
$d_{L}=0.81a$, $r=0.15a$, $\Delta=0.75a$ and $\varepsilon=13.0$.
The point defect is created by doubling the radius of a periodic waveguide
with period~$d_{L}$. Arrows mark the resonances discussed in the
text.\label{Fig:cantor}}
\end{figure}

In order to compare the amount of energy gathered by the coaxial terminal
and dissipated elsewhere, we excite the system by a dipole source
emitting a pulse with a broad spectrum, and use the FDTD method to
investigate the process of energy loss into the surroundings. 
Fig.~\ref{Fig:cantor}
shows the results. For the point-defect structure, the radiation of
the resonant mode primarily escapes sideways (Fig.~\ref{Fig:cantor}~(a)),
so despite having a high $Q$-factor ($2.1\times10^{4}$), the coupling
between the resonator and other components cannot be made efficient.
The Cantor structure shows a considerably improved coupling (Fig.~\ref{Fig:cantor}~(b))
accompanied by a drop of the $Q$-factor down to $2.7\times10^{3}$. Using
a W3 CNPW with Cantor geometry (Fig.~\ref{Fig:cantor}~(c)) raises
it back to $Q=1.1\div2.4\times10^{4}$ while still providing just
as good coupling to the coaxial terminal.

One should notice that the Cantor geometry is only one kind of 
deterministically aperiodic arrangement. Other kinds, e.g., 
quasi-periodic Fibonacci-like
one, can be used leading to a modification of the mode structure of
NPWs as well as the coupling efficiency of resonant mode into coaxial
terminal \cite{zhukJOSA}. Engineering the longitudinal geometry of
CNPWs appears to be a promising and powerful tool for a further degree
of freedom in controlling the dispersion properties.

\section{\label{Sec:Applications}Applications \textbf{}}

Relatively high transmission efficiency and flexibility in dispersion
tuning of CNPWs may initiate their use as components for efficient
and compact nanophotonic devices. Here we discuss two possible applications
of CNPWs in integrated optics: a coupled nanopillar waveguide directional
coupler \cite{ref20} and a switchable coupled mode laser \cite{zhukPSS}.

\subsection{Directional coupler}

\begin{figure}[t]
\begin{centering}\includegraphics[width=0.8\columnwidth]{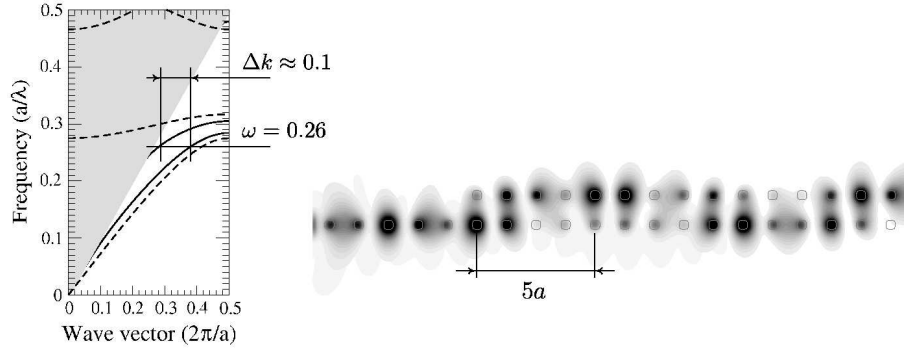}\par\end{centering}

\caption{\label{Fig:coupler}Dispersion diagram for the W2 CNPW section of a directional
coupler (left). Directional coupler based on two W1 CNPWs (right).
Grey levels mark field intensity. Here $\varepsilon=13.0$ and $r=0.15a$.}
\end{figure}

A pair of CNPWs can be used as an effective directional coupler \cite{ref20}.
An example of such a directional coupler based on two W1 waveguides
is shown in Fig.~\ref{Fig:coupler}. Analyzing the dispersion diagram
of the coupling section, namely W2 CNPW (Fig.~\ref{Fig:coupler},
left panel), one can see a pronounced difference in the propagation
constants of the even and odd supermodes in the frequency region around
$\omega=0.25-0.27$. It is a result of the strong interaction of coupled
waveguides, which now are much closer to each other than in the case
of standard line defect waveguides in a PhC lattice (see, for example,
the similar rod structure in \cite{ref21}). In this frequency range
the difference between the even and odd supermode propagation constant
is close to $0.1\cdot2\pi/a$, which leads to a crude estimate
of minimum coupling length \cite{ref22}: $L=\mathit{\pi}/|k_{even}-k_{odd}|=\left(\pi/0.1\right)\cdot\left(a/2\pi\right)=5a.$
Enhanced interaction leads to a shorter coupling length. This is
illustrated in figure~\ref{Fig:coupler}, right panel, where
the time averaged squared electric field pattern is shown for the
normalized frequency $\omega=0.26$. Guided light hops from the bottom
W1 waveguide to the top one and back on a distance equal to approximately
$5a$, which represents well the estimated value. In contrast to a
directional coupler proposed in \cite{ref23}, the CNPW structure does
not require a specially adjusted separation layer between coupled
waveguides, thus considerably simplifying a directional coupler design
and fabrication. 

There are several parameters, which can be used to optimize the directional
coupler, e.g., length of the coupling region or the number of rows in
each of the waveguides. As it has been shown in Sec.~\ref{Sec:Dispersion},
longitudinal and transverse offsets between the individual waveguides, 
as well as variation of the dielectric
constant and radius of the rods substantially modify the dispersion of
the compound system, thus affecting the coupling efficiency. For example, by
shortening the distance between two waveguides one can dramatically
increase the propagation constant difference and reduce the coupler size.
A similar effect cannot be achieved with standard PhC waveguides without
any special design tricks involving intermediate walls, which increases
the complexity of the fabrication procedure. An arbitrary longitudinal
offset breaks the symmetry of the device with respect to the symmetry
plane between the two W1 waveguides, which may further improve the coupling
strength similar to the case of an antisymmetric grating coupler \cite{ref24,ref25}.

\subsection{Laser resonators}

The periodicity of the coupled nanopillar waveguides can ensure the distributed
feedback within a finite waveguide section. This can be seen from the
flat tails of the nanopillar waveguide modes near the IBZ edge, which
correspond to a very low group velocity of CNPW modes. Taking into
account that any of the CNPW modes can be efficiently excited in the waveguide
using an external seeding signal of the appropriate spatial 
profile \cite{ref14}, we have proposed the
design of a switchable laser resonator \cite{zhukPRL} with distributed
feedback based on the CNPW. The possibility to tune the
number of modes, their frequency and separation (Sec.~\ref{Sec:Dispersion})
would make such a resonator a promising candidate for a chip-integrated
laser source.

The concept of switchable lasing was originally proposed in Ref.~\cite{ref14}
and has got further justifications in our recent work \cite{zhukPRL}.
In essence, a switchable microlaser comprised of a multimode microresonator,
where lasing can be switched on demand to any of its eigenmodes by
injection seeding \cite{Siegman,lee03}, i.e. by injecting an appropriate
pulse before and during the onset of lasing, such that the stimulated
emission builds up in a designated mode selected by this seeding field rather
than from the random noise present in the system due to quantum fluctuations
and spontaneous emission \cite{zhukPRL}.

To provide a basic physical picture of switchable lasing we first
consider briefly a simple semi-classical laser model in the case of
two identical coupled single-mode cavities \cite{zhukPRL}. In this
case there are two modes, the symmetric and the antisymmetric one,
characterized by spatial field distributions $u_{1,2}(\mathbf{r})$
and frequencies $\omega_{1,2}=\omega_{0}\mp\Delta\omega$, respectively.
Here $\Delta\omega$ is the mode detuning from the frequency of the
single-cavity resonance, $\omega_{0}$. For weak mode overlap the
spatial intensity profiles of the two modes nearly coincide, $\left|u_{1}(r)\right|^{2}\approx\left|u_{2}(r)\right|^{2}$.
We assume that the cavities contain a laser medium with a homogeneously
broadened gain line of width $\Delta\omega_{a}>\Delta\omega$, centered
at frequency~$\omega_{a}=\omega_{0}+\delta$. Here $\delta$
is the detuning of the gain profile from the cavity frequency $\omega_{0}$.
For this system the semiclassical Maxwell-Bloch equations \cite{Siegman,LambLaser},
in the rotating-wave and the slowly varying envelope approximation
read\begin{eqnarray}
\frac{dE_{1}(t)}{dt} & = & gR_{1}\left(\mathcal{L}_{1}-\frac{\kappa_{1}}{gR_{1}}\right)E_{1}(t)\nonumber \\
 & - & gR_{1}\eta\mathcal{L}_{1}\left(\alpha_{11}^{11}\mathcal{L}_{1}\left|E_{1}\right|^{2}+\left[\alpha_{22}^{11}\mathcal{L}_{2}-\alpha_{21}^{12}\mathrm{Re}\left(\chi_{1}\mathcal{M}_{12}\right)\right]\left|E_{2}\right|^{2}\right)E_{1}(t)+F_{1}(t),\nonumber \\
\frac{dE_{2}(t)}{dt} & = & gR_{2}\left(\mathcal{L}_{2}-\frac{\kappa_{2}}{gR_{2}}\right)E_{2}(t)\nonumber \\
 & - & gR_{2}\eta\mathcal{L}_{2}\left(\left[\alpha_{11}^{22}\mathcal{L}_{1}-\alpha_{12}^{21}\mathrm{Re}\left(\chi_{2}\mathcal{M}_{21}\right)\right]\left|E_{1}\right|^{2}+\alpha_{22}^{22}\mathcal{L}_{2}\left|E_{2}\right|^{2}\right)E_{2}(t)+F_{2}(t).\label{eq:anal_fields}\end{eqnarray}
Here all the spatial dependencies of the electric field and atomic
polarization were represented in the basis of the two cavity modes,
such that $E(\mathbf{r},t)=E_{1}(t)u_{1}(\mathbf{r})e^{-i\omega_{1}t}+E_{2}(t)u_{2}(\mathbf{r})e^{-i\omega_{2}t}$,
etc., and the atomic polarization was eliminated adiabatically \cite{Hodges,JohnBusch}.
$E_{j}(t)$ are slowly varying envelopes of two modes $j=1,2$. In
Eqs.~(\ref{eq:anal_fields}) the terms linear in $E_{j}(t)$ describe
stimulated emission driving, where the light-matter coupling constant 
is denoted by $g\simeq\sqrt{2\pi\omega_{0}d^{2}/\hbar}$,
the pumping rates projected onto the two resonator modes by 
$R_{j}=\int_{G}u_{j}^{*}(\mathbf{r})u_{j}(\mathbf{r})R(\mathbf{r})d\mathbf{r}$,
and the cavity mode decay rates by $\kappa_{j}$. Here $d$ is the
dipole moment of the atomic transition. The coefficients $\mathcal{L}_{j}=\mathrm{Re\,}\beta_{j}^{-1}$,
with $\beta_{1,2}=\Delta\omega_{a}/2+i\left(\delta\pm\Delta\omega\right)$,
account for the different mode-to-gain couplings due to asymmetrical
detuning of the atomic transition with respect to the resonator frequencies.
The terms cubic in $E_{j}(t)$ describe field saturation above the
lasing threshold, where $\eta=d^{2}/2\gamma_{\parallel}\hbar^{2}$
and the overlap integrals $\alpha_{kl}^{ij}=\int_{G}u_{i}^{*}(\mathbf{r})u_{j}(\mathbf{r})u_{k}^{*}(\mathbf{r})u_{l}(\mathbf{r})d\mathbf{r}$
are taken over the regions $G$ containing the gain medium. 
Here $\gamma_{\parallel}$ is the non-radiative decay rate. 
The frequency dependence of the cross-saturation
terms is given by $\mathcal{M}_{ij}=\beta_{i}^{-1}+\left(\beta_{j}^{*}\right)^{-1}$,
$i\neq j$. Since $\left|u_{1}(r)\right|^{2}\approx\left|u_{2}(r)\right|^{2}$
we can further assume that $\alpha_{jj}^{ii}=\alpha_{ji}^{ij}\equiv\alpha$,
$R_{1}=R_{2}=R$ and $\kappa_{1}=\kappa_{2}=\kappa$.

\begin{figure}[t]
\begin{centering}\includegraphics[width=0.8\columnwidth]{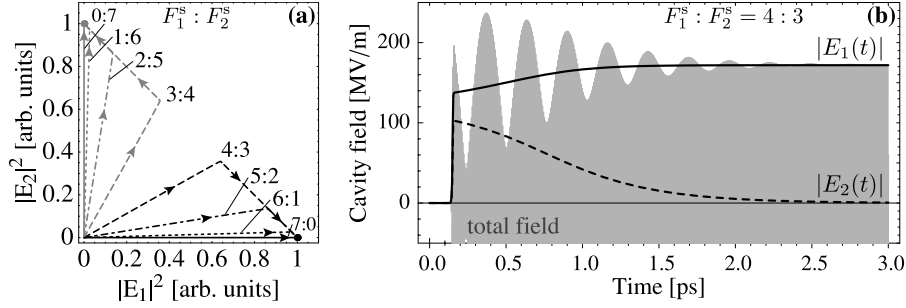} \par\end{centering}

\caption{\label{Fig:phaseDiagram}(Left) Cavity phase diagrams for a lasing
system governed by Eqs.~\eqref{eq:anal_fields} for $F_{j}^{\mathrm{s}}F(t)\gg F_{j}^{\mathrm{n}}(t)$
in the case of symmetric distribution of mode frequencies with respect
to gain. The dots denote the stable cavity states and the curves represent
the phase trajectories for their temporal evolution for different
ratios $F_{1}^{\mathrm{s}}:F_{2}^{\mathrm{s}}$ in the direction of
the arrows. (Right) Time evolution of the total laser field (shown
in grey) and the two cavity modes envelops $\left|E_{1}\right|$ (solid
line ) and $\left|E_{2}\right|$ (dashed line) for the ratio $F_{1}^{\mathrm{s}}:F_{2}^{\mathrm{s}}=4:3$.
One can see how the first mode wins the competition. Here numerical
values of the coefficients in Eqs.~\eqref{eq:anal_fields} were calculated
for W2 CNPW resonator with the following parameters, period $a$,
$d=1.21a,$ $r=0.15a$ and $\varepsilon=13.0$.}
\end{figure}

The inhomogeneous terms~$F_{j}(t)$ originate from the external injection
seeding field and from a noise field accounting for spontaneous emission
\cite{Hodges}. For vanishing functions $F_{j}$, Eqs.~\eqref{eq:anal_fields}
would take the form of the standard two-mode competition equations
\cite{Siegman,LambLaser}, describing bistable lasing \cite{BabaDisks}
and mode hopping in the presence of stochastic noise in the system
\cite{Bang}. If both an external seeding field $\mathcal{E}^{s}(\mathbf{r},t)$
and a stochastic noise field $\mathcal{E}^{n}(\mathbf{r},t)$ are
present in the cavity,
$\mathcal{E}(\mathbf{r},t)=\mathcal{E}^{s}(\mathbf{r},t)+\mathcal{E}^{n}(\mathbf{r},t)$, the
inhomogeneous terms are given by,\begin{equation}
F_{j}(t)\approx\frac{\omega_{j}\mathcal{L}_{j}}{\tau}\int_{t-\tau}^{t}dt'e^{i\omega_{j}t'}\int_{G}u_{j}(\mathbf{r})\mathcal{E}(\mathbf{r},t')d\mathbf{r}=F_{j}^{\mathrm{s}}F(t)+F_{j}^{\mathrm{n}}(t).\label{eq:anal_F}\end{equation}
The time integration in Eq.~(\ref{eq:anal_F}) is the averaging over
a time interval larger than $1/\Delta\omega$. The function~$F(t)$
is determined by the temporal dependence of the seeding signal $\mathcal{E}^{s}(\mathbf{r},t)$.
The coefficients $F_{j}^{\mathrm{s}}$~and~$F_{j}^{\mathrm{n}}(t)$
are determined by the spatial overlap of each mode with the seeding
and noise fields, respectively.

We consider the situation when the seeding prevails over the noise,
i.e., $F_{j}^{\mathrm{s}}F(t)\gg F_{j}^{\mathrm{n}}(t)$, before and
during the onset of lasing. After the onset the $E_{j}$ become so
large that the terms~$F_{j}$ have no effect anymore. In this situation
the evolution of the resonator will be determined by the ratio of
$F_{1}^{\mathrm{s}}$~and~$F_{2}^{\mathrm{s}}$. In Fig.~\ref{Fig:phaseDiagram}
(left) the phase trajectories of the temporal resonator state evolution
in the ($\left|E_{1}\right|^{2},\,\left|E_{2}\right|^{2}$) space
is presented for different values of $F_{1}^{\mathrm{s}}$~and~$F_{2}^{\mathrm{s}}$.
As seen in Fig.~\ref{Fig:phaseDiagram} (left), the lasing state
first reaches overall intensity saturation ($\left|E_{1}\right|^{2}+\left|E_{2}\right|^{2}=E_{s}^{2}$)
and then drifts towards one of the stable fixed points corresponding
to single-mode lasing (either $\left|E_{1}\right|^{2}=E_{s}^{2}$
or $\left|E_{2}\right|^{2}=E_{s}^{2}$). The drift happens on a longer
time scale than the initial overall intensity growth, and the intermode
beats decay fast after the lasing onset (Fig.~\ref{Fig:phaseDiagram},
right). The drift occurs towards the mode whose spatial and temporal
overlap with the seeding signal is larger, demonstrating a switchable
lasing behavior. It is important to note that even in the case of
asymmetric detuning of the cavity modes with respect to the gain frequency
($\delta\neq0$), single-mode lasing is achieved into the mode whose
spatial overlap with the seeding field, $F_{j}^{\mathrm{s}}$, is
largest, i.e., if one of the following conditions, $F_{1}^{\mathrm{s}}\gg F_{2}^{\mathrm{s}}$
or $F_{1}^{\mathrm{s}}\ll F_{2}^{\mathrm{s}}$, is satisfied.

\begin{figure}[t]
\begin{centering}\includegraphics[width=0.7\columnwidth]{Figs/fig12a}\medskip{}
\par\end{centering}

\begin{centering}\includegraphics[width=0.9\columnwidth]{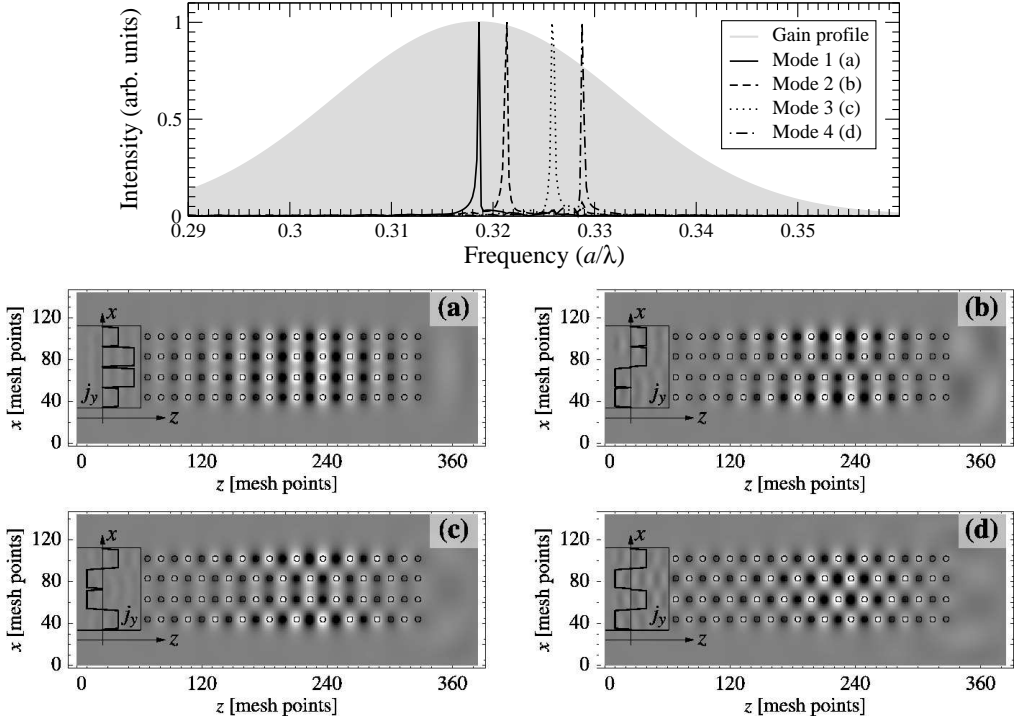}\par\end{centering}

\caption{The amplitude spectra (top) and laser filed distribution (bottom)
for the periodic injection-seeded four-row CNPWs. The lines labeled
{}``Mode-1'' to {}``Mode-4'' correspond to the seeding signals
(a)--(d) shown in bottom panel. The shaded areas represent the laser
amplification line, with its central frequency~$\omega_{a}=0.3225$.
The pumping rate equals $W_{p}=1.0\times10^{13}\,\textnormal{s}^{-1}$.
The panels (a)--(d) correspond to different seeding signals, shown
schematically as excited in the terminal. \label{Fig:periodicLasing}}
\end{figure}

To demonstrate the predictions of this simple theory we have modeled
the lasing action in four-row CNPW structure with a realistic injection
seeding (Fig.~\ref{Fig:periodicLasing}) using the FDTD method \cite{zhukPSS}.
The externally pumped laser-active medium is placed in the central
7 pillars of all four rows. This is done to maximize coupling between
the active medium and the main localization region of the lasing modes.
The population dynamics of an active medium is described at each space
point by the rate equations of a four-level laser with an external
pumping rate $W_{p}$. To achieve population inversion we have chosen
the following values for the non-radiative transition times, $\tau_{32}\simeq\tau_{10}\ll\tau_{21}$,
with $\tau_{31}=\tau_{10}=1\times10^{-13}\,\textrm{s}$, $\tau_{21}=3\times10^{-10}\,\textrm{s}$,
and the total level population is $N_{\textrm{total}}=10^{24}$ per
unit cell \cite{lasPauli}. The Maxwell equations are solved using
FDTD scheme supplemented by the usual equation of motion for the polarization
density in the medium and by the laser rate equations \cite{lasIEEE,lasRandom,lasPauli,lasJoannop}.
All calculations were done for TM polarization. The seeding signal
is excited by four emitters (linear groups of dipoles) engineered
on the regular dielectric waveguide attached to the CNPW structure
(see Fig.~\ref{Fig:periodicLasing}). Each of the emitters generates
a single short Gaussian pulse with carrier frequency $\omega_{a}$
and with half-width duration $\sigma_{t}=10^{4}dt$. The relative
phase of the fields in these pulses is chosen $0$ or $\pi$. Technically,
the seeding dipoles are realized as point like oscillating current
sources in the Maxwell equations \cite{zhukPSS}. Similarly, the spontaneous
emission \cite{lasRandom,lasSeo,Langevin} can be modeled as an ensemble
of point current sources, randomly placed in space, with temporally
$\delta$-correlated Langevin noise \cite{Langevin}. The computational
domain of size $7a\times22a$ was discretized with a mesh point spacing
of $a/16$. The time step is related to the spatial mesh to assure
stability and was chosen $dt=6\times10^{-17}\,\textrm{s}$. To simulate
an open system, perfectly matched layer (PML) boundary conditions
\cite{ref18} were used.

In Fig.~\ref{Fig:periodicLasing} (top) the lasing spectra in the
steady state long after the seeding signal has decayed is shown. The
broad shaded areas depict the laser line of width $\Delta\omega_{a}$
centered at $\omega_{a}$, which is shifted slightly towards lower
frequencies. As a rule the $Q$-factor is larger for modes with the
higher frequency. The shifted laser line compensates this $Q$-factor
difference, so that any of the four CNPW modes can be selected by
the appropriate seeding signal with the same symmetry. In figure~\ref{Fig:periodicLasing}
(bottom) the spatial electric field distribution in the four-row CNPW
laser resonator is shown at an instant of time long after the seeding
signal has decayed and after the steady state has been reached. The
symmetry of the selected lasing modes corresponds to that of the seeding
signal (Fig.~\ref{Fig:periodicLasing}). 

The proposed concept of switchable lasing is not limited to the periodic
CNPW structures, but is expected to work in any resonator featuring
bi- or multistability. Any coupled cavity based system would be a
good candidate for the effects predicted. For example aperiodic CNPW
based resonator also show the switchable lasing behavior for resonant
modes discussed in Sec.~\ref{Sec:QNPW} \cite{zhukPSS}.

\section{\label{Sec:Conclusion}Conclusion}

We have shown that a novel type of coupled nanopillar waveguides,
comprised of several periodic or aperiodic rows of dielectric rods,
may have potential applications in compact photonics. The strong coupling
regime can be utilized in ultrashort directional couplers or laser
cavities, which might possess an additional functionality and flexibility
when different longitudinal and transverse offsets among individual
waveguides are employed. The factors of major influence upon the mode
dispersion have been analyzed. Transmission spectra for 2D and 3D
systems prove the possible single mode excitation by imposing specific
symmetry conditions onto a field source and high transmission characteristics
of coupled nanopillar waveguides.

\begin{acknowledgement}
DNC, SVZ and JK acknowledge partial support by the Deutsche Forschungsgemeinschaft
through projects SPP 1113 and FOR 557. AVL acknowledges partial support
by Danish Technical Research Council via PIPE project and the EU Commission
FP6 via project NewTon (NMP4-CT-2005-017160).
\end{acknowledgement}

\end{document}